\documentclass[journal]{IEEEtran}
\IEEEoverridecommandlockouts
\usepackage{cite}
\usepackage{setspace}
\renewcommand{\baselinestretch}{1} \normalsize
\usepackage{amsmath,amssymb,amsfonts}
\usepackage{algorithmic}
\usepackage{graphicx}
\usepackage{textcomp}
\usepackage{xcolor}
\usepackage{url}
\usepackage{soul}
\usepackage{ulem}
\def\BibTeX{{\rm B\kern-.05em{\sc i\kern-.025em b}\kern-.08em
    T\kern-.1667em\lower.7ex\hbox{E}\kern-.125emX}}
\begin{document}

\title{\LARGE{deController: A Web3 Native Cyberspace Infrastructure Perspective}
}

\author{Hao Xu, Yunqing Sun, Zihao Li, Yao Sun, Lei~Zhang and Xiaoshuai Zhang  
	\thanks{This paper is submitted to IEEE for potential publication and it might be removed without notices. Corresponding authors: Xiaoshuai Zhang (main) and Lei Zhang.}
	\thanks{
		 Hao Xu E-mail: hxgla@outlook.com. Yunqing Sun is with Department of Computer Science, McCormick School of Engineering and Applied Science, Northwestern University, Evanston, IL, US, E-mail: yunqing.sun@northwestern.edu. Zihao Li, Yao Sun, Lei Zhang and Xiaoshuai Zhang are with University of Glasgow, Glasgow, G12 8QQ, UK, E-mail:  z.li.6@research.gla.ac.uk; \{Yao.Sun; Lei.Zhang; Xiaoshuai.Zhang\}@glasgow.ac.uk. 
		
	}}
\maketitle

\begin{abstract}
Web3 brings an emerging outlook for the value of decentralization, boosting the decentralized infrastructure. People can benefit from Web3, facilitated by the advances in distributed ledger technology, to read, write and own web content, services and applications more freely without revealing their real identities. Although the features and merits of Web3 have been widely discussed, the network architecture of Web3 and how to achieve complete decentralization considering law compliance in Web3 are still unclear. Here, we propose a perspective of Web3 architecture, deController, consisting of underlay and overlay network as Web3 infrastructures to underpin services and applications. The functions of underlay and overlay and their interactions are illustrated. Meanwhile, the security and privacy of Web3 are analyzed based on a novel design of three-tier identities cooperating with deController. Furthermore, the impacts of laws on privacy and cyber sovereignty to achieve Web3 are discussed. 
\end{abstract}

\begin{IEEEkeywords}
Web3 architecture, overlay and underlay, decentralized infrastructure, blockchain, DAO
\end{IEEEkeywords}
\section{Introduction}

\begin{figure}[tbp]
    \centering
    \includegraphics[width =0.47\textwidth]{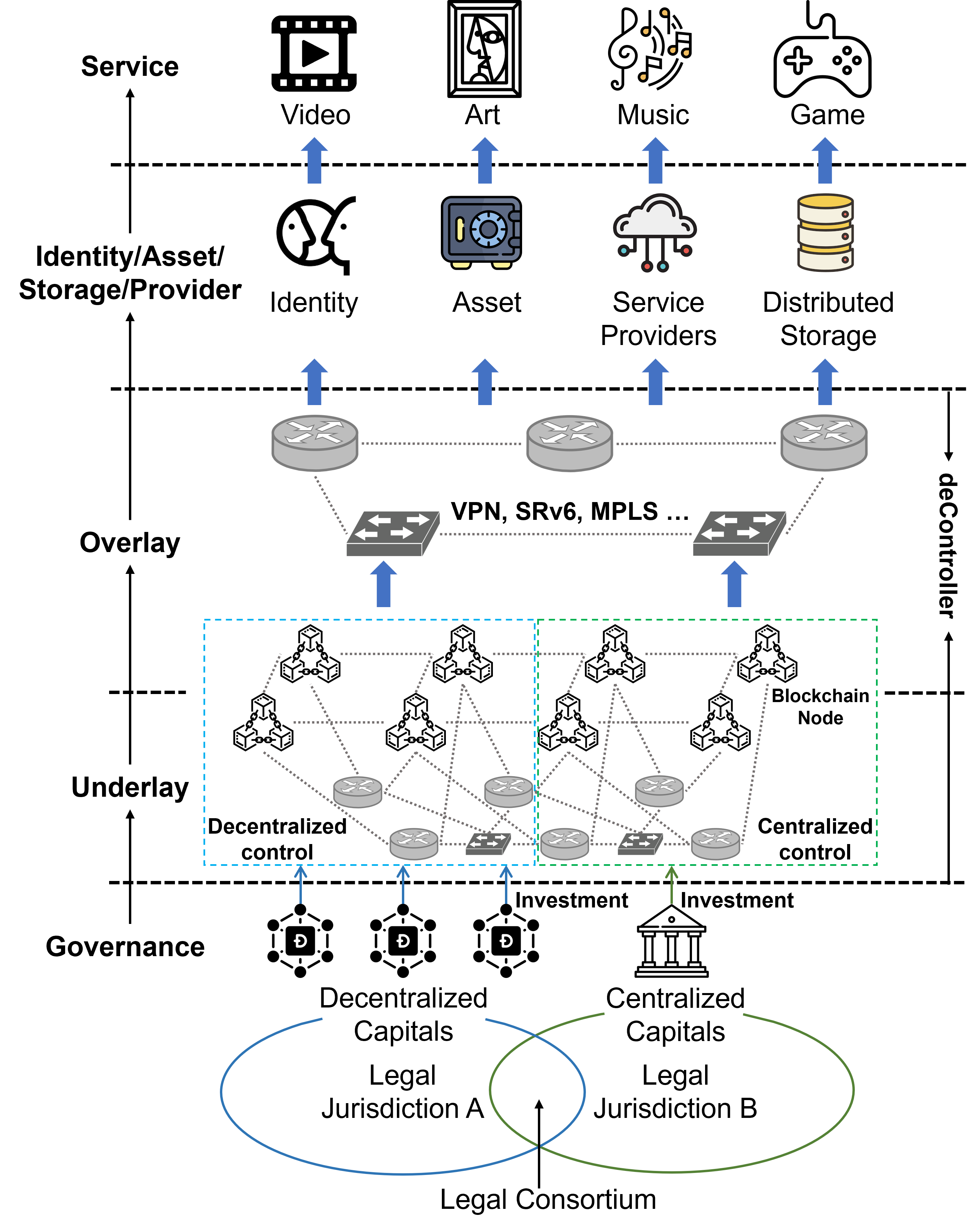} 
    \caption{Our contribution to the Web3 network architecture.}
    \label{fig:novelty}
\end{figure}

Web3, an emerging term of decentralized world-wide-web (WWW) based on distributed ledger technology (DLT) and crypto economy, has been foreseen as a dynamic-driven factor for the next generation of the Internet. Web3 is seen as a catalyst for the future Internet to provide content, services, and applications for users without centralized servers. Since the introduction of blockchain by Bitcoin in 2008, the decentralized network started its unprecedented journey and has been thriving for more than a decade. With the advances of blockchain, cryptocurrencies and decentralized autonomous organizations (DAO) have shifted the world to embrace the value of decentralization to deconstruct the well-established centralized WWW ecosystems with the decentralized governance, underlay and overlay network infrastructures, as shown in Fig. \ref{fig:novelty}, detailed in the following sections.

Currently, Web3 has reached the moment towards inclusive top-down solutions and the soil for its growth in industrial, commercial and public networks without the involvement of any centralized things, solidifying the lifeline of Web3 value and consensus. However, such a top-down architecture of Web3 has not been sculptured with considering its challenges as well as the interactions of network infrastructure, DLT, security and privacy, judicature, etc., comprehensively.

\subsection{Challenge and opportunity}
With a great boost on security, privacy and cyber sovereignty ({\color{black}cyber sovereignty refers to the cyber boundary established by a country or region for exercising national control and implementing specific legislation}) of user data, the challenges faced in achieving Web3 and opportunities are significant. 

\subsubsection{Web3 is running on centralized things!}
``Read, write and own" endorses the fundamental value in Web3; however, if the access to the space of Web3 is denied, ownership means little or nothing to the owner who is blocked from accessing the WWW. Meanwhile, the value of privacy offered by Web3 becomes void if the user can be tracked at the beginning and the end of Internet access. It is necessary to ensure the user will never be unplugged from the network or illegally tracked due to centralization causes. Most importantly, Web3 shall secure itself from running the whole network on the infrastructure offered by centralized resource controllers. 

Another distinct challenge is the authentication in access control of Web3 because all identities in the decentralized network are anonymous, i.e., the authentication should not reveal any personal information of Web3 users. However, the authentication information of users is known by the central controller in the centralized model. Therefore, the architecture to achieve anonymous authentication for decentralized Web3 should be further investigated.
\begin{figure}[htbp]
    \centering
    \includegraphics[width =0.47\textwidth]{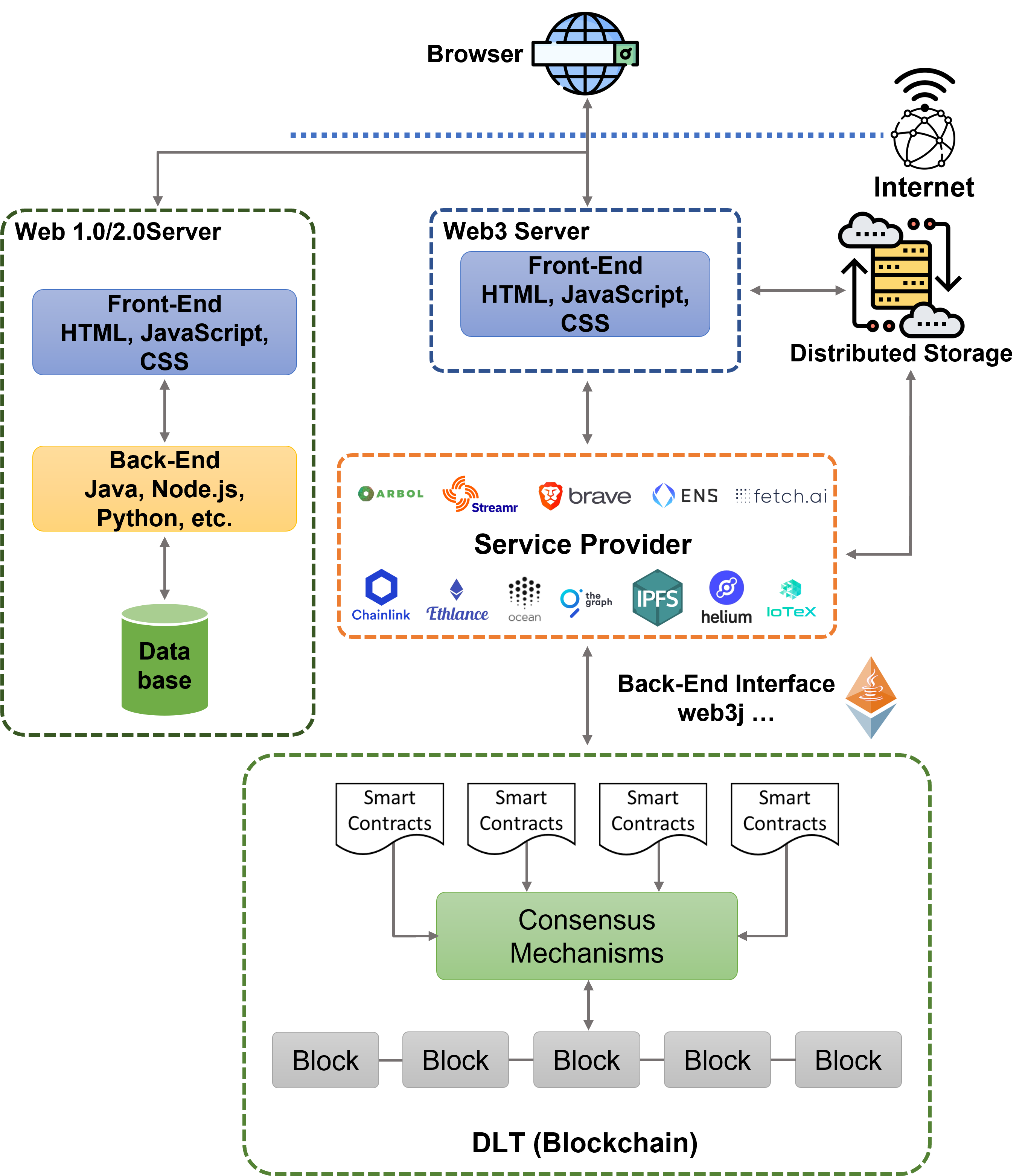}
    \caption{Architecture overviews of Web3 and Web 1.0/2.0.}
    \label{fig:web3-arc}
\end{figure}
\subsubsection{Opportunities}
Since the current web structure is highly centralized, Web3 could facilitate the shift from centralized Internet to decentralized Internet based on DLT, distributed network, NEAT (Network Encrypted Address Translation, detailed later), etc. Such a self-governance evolution may enable people to access and own Internet resources more freely and equally, boosting investments in Web3 network infrastructure, and owning the actual Web3 network. 
Privacy is also an opportunity as anonymity may challenge the legislation and jurisdiction. As a nature of Web3, anonymous identities can protect users' real identities to avoid censorship when users are involved in various activities and applications. It is inspiring to enable a fully private and connected universe for all via encrypted address, a.k.a. $\sf{BCADD}$, and the encrypted infrastructure through the ideology of decentralized and encrypted infrastructure. In this case, anyone who onboards the Web3 network can have permissionless access to the Web3 infrastructure. 


Apart from technological innovations, Web3 has the potential to provide new opportunities for legal governance of cyberspace due to its privacy-driven design. The core privacy issue in Web 1.0/2.0 is centralized services since service providers may exploit the surplus of online content creators without permission infringing users’ privacy and data protection rights, and even act as unsupervised police. Web3, embedded a native decentralization and encryption, offers users more control over their personal data and privacy in the Internet access where they own their autonomy to make choices, which is essentially aligned with the objectives of GDPR (General Data Protection Regulation) in the EU.


\subsection{Motivation and contribution}
There will be emerging scenarios relying on decentralization as its core value. Hence, it is necessary to prepare the existing network, security and privacy infrastructure to embrace the world of decentralization, meaning the infrastructure as a whole needs to stand with decentralization value rather than an unavoidable connection with centralization. Therefore, we propose the enabling decentralization infrastructure controller, deController, for Web3 native infrastructure. 

This paper contributes to Web3 in three aspects: (a). the Web3 network architecture with the detailed description of deController consisting of the overlay and underlay network; (b). the security, privacy and identity in a fully decentralized manner; (c). the operational principles regarding law and governance for Web3 infrastructure as shown in Fig. \ref{fig:novelty}.

\section{Web3 Outlook in network and services}
Compared with the centralized network, such a decentralized network structure brings different considerations in Web3 such as where the data are stored, how to ensure the data validity, etc. {\color{black}On the other hand, existing peer-to-peer routing and network protocols, such as Chord and Distributed Hash Table (DHT) can enable overlay connectivity.}

\subsection{Web3 architecture overview}
The network architecture of Web3 is depicted in Fig. \ref{fig:web3-arc}. Compared with the network architecture of Web 1.0/2.0 using a centralized web server to provide web services as shown in the left of Fig. \ref{fig:web3-arc}, the Web3 server runs in a more decentralized manner. Specifically, the Web3 server only provides frontends of services while data storage and backends of applications are provided in a distributed manner. {\color{black}Users can access an application via the blockchain address of the corresponding smart contracts, in which the application backend is contained. Blockchain addresses can be routed by the Web3 network in accessing the application. The data content of users and applications (images, voice, videos, etc.) may be stored in a distributed storage to avoid data corruption or loss. Lawful agreements on access control policies can be applied to user data stored by service providers.} 


{\color{black}To protect the real identities of users, a 3-tier identity architecture is proposed in Section \ref{three-tier} to avoid personal information leakage and identity tracing.} {\color{black}In addition, the data of identity mapping to the network and transactions between users and applications, such as payments and records of purchased items/services, can be recorded by DLT in public ledgers as they are small data compared to the content data.} Such records can only be written into public ledgers after being verified by consensus mechanisms in the Web3 network, so the records are transparent, undeniable and immutable. Therefore, users and service providers cannot forge records or distort the existing records. Even if applications are shut down by service providers, users' assets in applications are kept in public ledgers, where users can access their assets seamlessly at their own discretion. Such a feature is difficult to be natively supported by applications in Web 1.0/2.0 since user data is fully controlled by service providers in centralized servers. 

\subsection{Overlay and underlay decentralization of Web3 network }\label{three-tier}
The decentralization of network has never been easier with the help of blockchain. Regardless of the consensus type, each blockchain full node operates a full stack of networking and servicing protocols, making them a perfect nexus for the decentralized network. In fact, the existing blockchain delivery network makes the suitable alternative for the Web3 overlay network, as shown in Fig. \ref{fig:novelty}. 
{\color{black}The underlay can be regarded as the 5-layer common computer network to provide physical network connections for the overlay. NEAT is used to resolve the association of $\sf{BCADD}$ with any network device identifiers, network ports and domain names. By linking the $\sf{BCADD}$ to specific identifiers, deController is able to lookup the $\sf{BCADD}$ globally and establishes the overlay network in any given underlay network.} With the underlay network used by the blockchain delivery network, the underlay network will grow in the decentralization's interest, hence becoming the decentralized underlay network operated under the principle of fully decentralized infrastructure, which is illustrated in detail later in Section \ref{sec:operate}.

In the Web3 context, the role of underlay network nodes overlaps with {\color{black}the blockchain nodes in the overlay owned by different stakeholders such as companies and organizations.} These nodes also play the pivotal role of supplying computing power and networking capacity of the blockchain network. In fact, blockchain nodes can also provide the necessary overlay tunneling and routing capabilities, hence becoming the pillar of the Web3 overlay network.

\subsubsection{Decentralized Applications and Services}
The Web3 features the owner economy, which boosts the decentralized applications (dApp). The dApp is a smart contract powered autonomous code running on decentralized networks. Once the code is deployed on the blockchain, it becomes a public asset for
any entities within the network. However, the dApp only works as an agent passing on the value between users; it cannot offer demanding services, e.g., video streaming, chatting room or online gaming. To enrich the context of Web3 ecosystem, service providers can use dApp to securely provide services to users using encrypted identities and exchange tokens, hence becoming a decentralized service provider. 

\subsubsection{Decentralized Network Infrastructure}

In the scope of network infrastructure, the aforementioned Web3 network architecture is logically divided into two layers, the underlay and the overlay, as shown in Fig. \ref{fig:novelty}. Similar to the traditional network, the underlay in Web3 architecture can be divided into multiple segments, which are later tagged by the overlay blockchain node with the optimal topological resolution. The entities within each segment perform particular network functions in a decentralized way, which is critically different from traditional networks. As mentioned, two decentralization manners, P2P (Peer-to-Peer) and DAO2DAO \cite{dao2dao} (federated), can be exploited in underlay depending on the function performed. For example, multiple computing servers organized by a DAO in the edge network segment can provide route optimization service for the Web3 overlay network, collaborating with different DAOs in a decentralized manner, while the entities' data flow can be organized in a P2P manner that matches an optimal route offered by the Web3 overlay network, in order to manage the packet delivery for users.   

The overlay is built above the underlay to control and manage this decentralized network in the Web3 architecture \cite{Xu2021beran}. Generally, the main entity in the overlay is the controller in charge of all the network management functions including authentication (identity and access), data packet routing, computing resource allocation, etc. These functions will be elaborated in Section \ref{reshaping}.

\subsubsection{Integration: An identity prospect of view}
Since Web3 aims for a decentralized network where users can control their data and identity revocation or reservation, most user identities are self-sovereign identities {\color{black}(self-sovereign refers to empowering users to control their own identity information in cyberspace)} rather than centralized or federated identities. However, a hierarchy and decentralized identity management infrastructure is necessary to construct a uniform identity authentication scheme that crosses the different worlds and domains.

\begin{figure}
    \centering
    \includegraphics[width =0.5\textwidth]{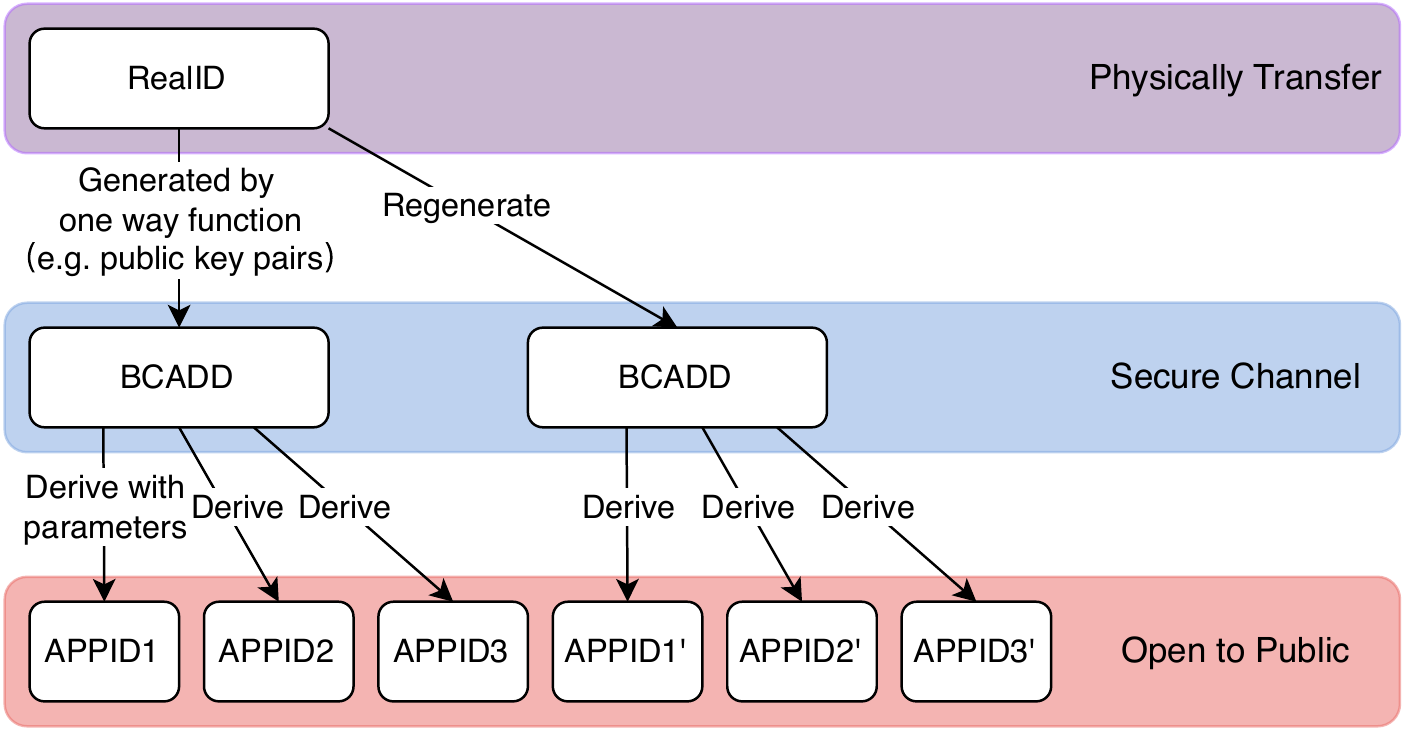} 
    \caption{Three tiers of identity.}
    \label{fig:identity}
\end{figure}

{\color{black}A 3-tier identity management is proposed to bridge the real identity to virtual identity from the perspective of users and services in Fig. \ref{fig:identity}. A real user identity can be linked to several virtual user identities to represent the user in Web3 networks. Meanwhile, a virtual user identity can derive the identities of multiple applications and services since a user may operate different applications and services.} Therefore, a user's identity in the real world is regarded as the first level identity, named ${\sf{RealID}}$. ${\sf{RealID}}$ is confidential and never revealed in Web3. The second level identity is the address of the user's wallet, which is called ${\sf{BCADD}}$ \cite{Xu2021beran}. ${\sf{BCADD}}$ is derived from ${\sf{RealID}}$ locally by a one-way function and used by network operators. The third level identity is regarded as the application ID, 
which is used as the identity in different services, named ${\sf{APPID}}$. ${\sf{APPID}}$ is derived from the current ${\sf{BCADD}}$ together with properties of the service by one-way function or verified random function (VRF) \cite{bitansky2020verifiable}. The ${\sf{APPID}}$s are the self-sovereign identities in the applications, end-to-end routing, and services of Web3 but both ${\sf{BCADD}}$s and ${\sf{APPID}}$s are hardly being traced without the parameters of the one-way function. Since the overlay network hosted by blockchain networks can lookup every $\sf{BCADDs}$ in a global view and route every traffic between them, the direct connections between two encrypted identities can be established. Hence, the user can use a unified address authentication based on the public-key identity.

{\color{black}{As the decentralization, anonymity and privacy are the prominent factors of Web3, where public keys are used as identities, CAs (certificate authorities) are not required for the authenticity endorsement of identities and ownership of public keys.}} {\color{black}{However, the use of public-key-based identity poses a security threat to the regulatory management of citizen networks, as they are fully anonymous and self-issued. It is a challenge to obtain the real identity of users without knowing a prior identity association to the public-key-based address. Therefore, the regulator should require mandatory registrations of active public-key-based identities to comply with the regulation. On the other hand, the legal interception can also be implemented into the deController through steering and duplication of traffics.}}

Here, we first define the visibility of the three layers' identities followed by their security levels. ${\sf{RealID}}$ is only held by the user and registered at the regulatory body where necessary. ${\sf{BCADD}}$ is public to the cryptocurrency system and other authorized infrastructure, including the Mobile Network Operator (MNO) and Internet Service Providers (ISP). 
When the ${\sf{BCADD}}$ is derived from the ${\sf{RealID}}$, the user can decide to involve more information in the ${\sf{BCADD}}$ using VRF and zero-knowledge proof (ZKP) \cite{zeestar} via the regulatory body. {\color{black}{As required in Web3, personal information could not be revealed in the network. However, when any information is needed in the network or applications, the regulatory body can apply zero-knowledge proof on the user's registered information and publish a proof to the network. In this way, the user can prove to the Web3 application that it has the information or attribute as required.}}
For example, a user can state its age is over 18 without revealing the actual age using ZKP and VRF in the statements linked to ${\sf{BCADD}}$. ZKP and VRF enable service providers to verify the authenticity of the statement using the ${\sf{BCADD}}$. 
In addition, ${\sf{APPID}}$ is visible to any service providers on Web3. To resist tracking attacks and protect users’ contextual privacy, the identity information should be updated in the user-defined privacy time slot. 
To be consistent with regular updates of wallet addresses, once the ${\sf{BCADD}}$ is updated, the ${\sf{APPID}}$ should also be updated at the service provider. 
In this case, the ${\sf{APPID}}$ of the same user may be linked to different wallets to impair the service consistency and interruptions. 

By having the 3-tier identity hierarchy in Fig. \ref{fig:identity}, it is ensured that different identities in different domains with different security levels. 
${\sf{BCADD}}$ should be known to the operator to determine if a subscription for the network access is valid. The ${\sf{APPID}}$ is used as both the network interface indicator and the service authentication account. Since ${\sf{APPID}}$ is derived from ${\sf{BCADD}}$, the authentication of ${\sf{APPID}}$ by the AAA (Authentication, Authorization, and Accounting) server can be done over the distributed network provider when the routers and RAN  {\color{black}(Radio Access Network)} authenticate and trust the ${\sf{BCADD}}$.   
The detailed architecture of ${\sf{BCADD}}$ in-network accessing and ${\sf{APPID}}$ in application service with security and privacy authentication are shown in Fig. \ref{fig:sparch} and described as follows.

\subsection{Security and privacy }
In Web3 infrastructure, all the blockchain nodes and user identities should be registered and updated to the blockchain platform by sending a bootstrapped transaction. 
{\color{black}The transaction should include the blockchain node’s ${\sf{BCADD}}$ to support network service, or ${\sf{APPID}}$ together with the access control information to support application services.} The access control information may include a Non-Fungible Token (NFT) or other legacy server addresses. {\color{black}{NFT has been recognized as a unique identifier of key digital assets, so the possession of certain assets represents the privilege of objects in the form of ownership in a manner of attribute-based access. Such adoption of the ownership concept can be migrated into access control aspects, where the ownership represents the access privilege.}} After registration, once the blockchain node requires a service from a third-party application server (AS), it will initiate a decentralized mutual authentication\cite{Xu2021beran} of ${\sf{APPID}}$ between them. To be compatible with the protocol in \cite{Xu2021beran}, we let 
all the routers check the ${\sf{APPID}}$s directly and pass the packets transparently to continue the authentication between users and AS. After the authentication procedures are finished, the AS will check the authenticated ${\sf{APPID}}$’s corresponding authority by searching the access control information in the blockchain platform. When the ${\sf{APPID}}$s are renewed together with ${\sf{BCADD}}$, users can decide to keep service consistency by notifying the new ${\sf{APPID}}$ in the previous session or the old ${\sf{APPID}}$ in the new session.
\begin{figure}[tbp]
    \centering
    \includegraphics[width = 0.5\textwidth]{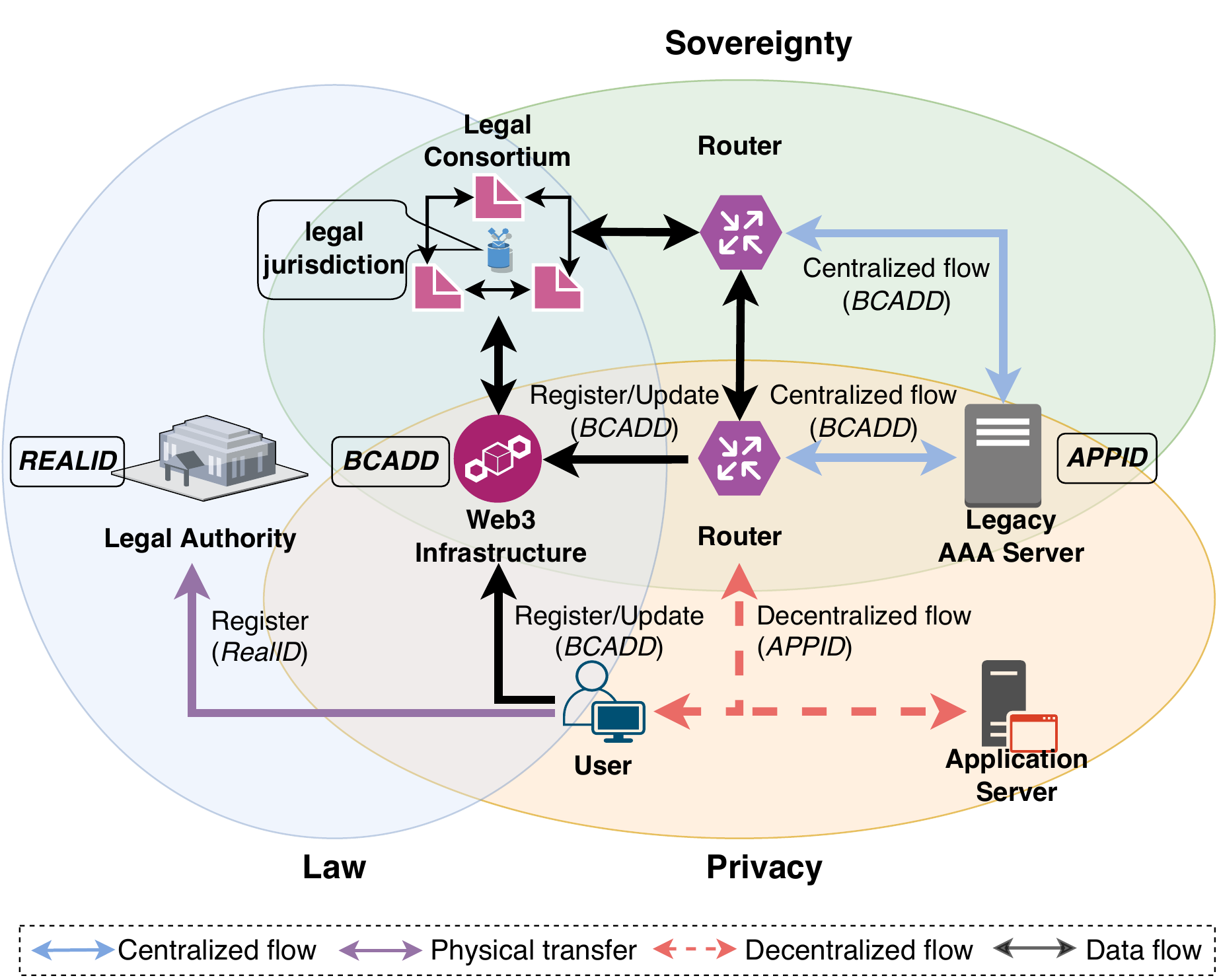}
    \caption{Security and Privacy architecture.}
    \label{fig:sparch}
\end{figure}
The checking procedure executed by the first router is as follows. Authentication between the user and the communications network is first implemented to authenticate {\sf BCADD} and support communications. By the derivation relation between ${\sf{BCADD}}$ and ${\sf{APPID}}$, the ${\sf{APPID}}$ can be verified and authenticated given ${\sf{BCADD}}$. Once the user can prove to the router that the ${\sf{APPID}}$’s holder has a valid ${\sf{BCADD}}$ and has initiated a valid transaction for this session without revealing any information, the first router will forward the message to the destination. 


The traditional AAA server still exists in Web3 in case decentralized authentication is incompatible with any third-party AS or user. A legacy registered AAA server can run the General Bootstrapping Architecture (GBA) protocol \cite{33.220} to generate a secure channel between the user and AS. The legacy AAA server should also register to the blockchain platform to be compatible with the blockchain network infrastructures, as shown in Fig. \ref{fig:sparch}. The blockchain in the Web3 network can be regarded as a random oracle to execute computation under public supervision. Meanwhile, new privacy-preserving techniques, like public verifiable ZKP, can be introduced and implemented in the blockchain platform to provide Web3 transparent and regulated privacy protection. 

\section{DAO for decentralized communication infrastructure: Reshaping the underlay network}\label{sec:operate}
As aforementioned, decentralization is identified as the core interest of Web3, led by the blockchain (DLT), dApp, DeFi, and DAO. Communities of decentralization have become the beneficiary of decentralized networks, regardless of the fact that the current whole network is built on top of centralized communication infrastructures. However, dApps cannot be considered fully decentralized with their roots in centralized infrastructures. Therefore, there is a requirement that the underlay of the whole decentralized network, namely communication infrastructure operators, become decentralized.

\subsection{Motivation of DAO-based infrastructure operator} 
With the requirement of full decentralization, DAO has the potential to fully decentralize the infrastructure operator. Unlike the traditional telecommunication business entities (e.g., state-owned, private-owned, public-limited, and limited liability companies), the DAO-based infrastructure operator has decentralized structures in essence. Firstly, the DAO-based infrastructure operator can flatten out the entire corporate management structure. There is no centralized management role to really control the organization. Instead, the vital decision can be proposed and made by every member of the organization, namely, DAO stakeholders.

Secondly, the organization’s rules are encoded using innovative contract technology in a permissionless blockchain. {\color{black}Traditional organizations do not have to maintain complex and costly administrative departments.} DAOs also make it virtually impossible to commit fraud since every transaction is open to public and consortium scrutiny. Another feature of a DAO is that the decisions are executed automatically via votes on the blockchain using smart contracts, which are transparent and non-repudiate. Once a proposal has been successfully voted upon, change occurs automatically without the need for further human involvement.

DAOs represent a radical rethink of how infrastructure can be structured and operated, including changes in ownership, governance, decision-making and profit distribution. Decentralized infrastructure operators can not only inspire the investment in Web3 infrastructures, but also reshape legal consortium through the use of smart contracts, as shown in Fig. \ref{fig:novelty}. With the demand of full decentralization, DAO could extend to telecommunication infrastructural operators \cite{Xu2022metaverse}, operating the entire underlay and overlay network nodes with its own natural resources, such as spectrum, computing resources and energy. {\color{black}Furthermore, DAOs are always motivated to add more value to their content and services created in Web3. However, the value based on decentralization and consensus cannot be secured if the underlay network and storage are built upon the centralized infrastructure. Therefore, another major motivation for DAOs to invest in decentralized infrastructures is to protect their key assets in Web3, while making communal profits from serving Web3 people in the future.}
Although DAO-based infrastructure operators have many advantages, one of the biggest challenges is the risks of legal compliance to cyber sovereignty and data protection law when infrastructure operators become decentralized and multinational.

\subsection{A legal view on decentralized infrastructure of Web3}


As illustrated in Fig. \ref{fig:sparch}, the decentralized underlay of Web3 significantly impacts law, privacy and cyber sovereignty. Although the decentralized infrastructure has the potential to address the cybersecurity and sovereignty risks associated with data cross-board flow, there are still some potential frictions between the decentralized underlay and the current legal system.


Firstly, full anonymization is still difficult to be achieved since operators or governments are possible to retrieve personal data through a combination of data from network activities even though the 3-tier identity is applied. However, the recovery possibility is also necessitated by the government's legitimate surveillance requirements, which can be a tool for cyberspace regulation. In such a scenario, the government should define anonymization in data protection laws clearly \cite{Finck2019} and obey the purpose limitation principle by complementing legislation to mitigate risks.

The second potential friction is users can actually own {\color{black}partial} Web3 network and contribute to it under the decentralized infrastructure. However, they may make themselves ``network operators'' or ``data processors'' within the meaning of the Cybersecurity Law or Data Protection Law (e.g., GDPR). Thus, they theoretically have to bear the corresponding legal responsibilities for data protection.
Such a design does not fully consider the challenges posed by decentralization and the decentralized infrastructure. Therefore, it leads to the critical reflection of regulatory philosophy in this decentralized privacy-friendly architecture, which requires a new legal paradigm of cyberspace regulation. 
Government-led regulatory impact sandboxes could act as stabilizers {\color{black}to calibrate the law and technology} for industry compliance, maximizing the compatibility of Web3 with existing legal systems and regulatory regimes.

Moreover, the decentralized infrastructure may result in data flowing to different jurisdictions, indicated in Fig. \ref{fig:sparch}, creating jurisdictional conflicts and jeopardizing national cybersecurity and sovereignty. The legal consortium introduced in Fig. \ref{fig:novelty} may enable different jurisdictions to reach a consensus by smart contracts on issues of judicial jurisdiction. Therefore, it may eliminate unauthorized cross-border movements of data and ensure national cybersecurity and sovereignty.

\section{The Web3 deController for Network infrastructure: Reshaping the overlay network}\label{reshaping}
\begin{figure}
    \centering
    \includegraphics[width=0.51\textwidth]{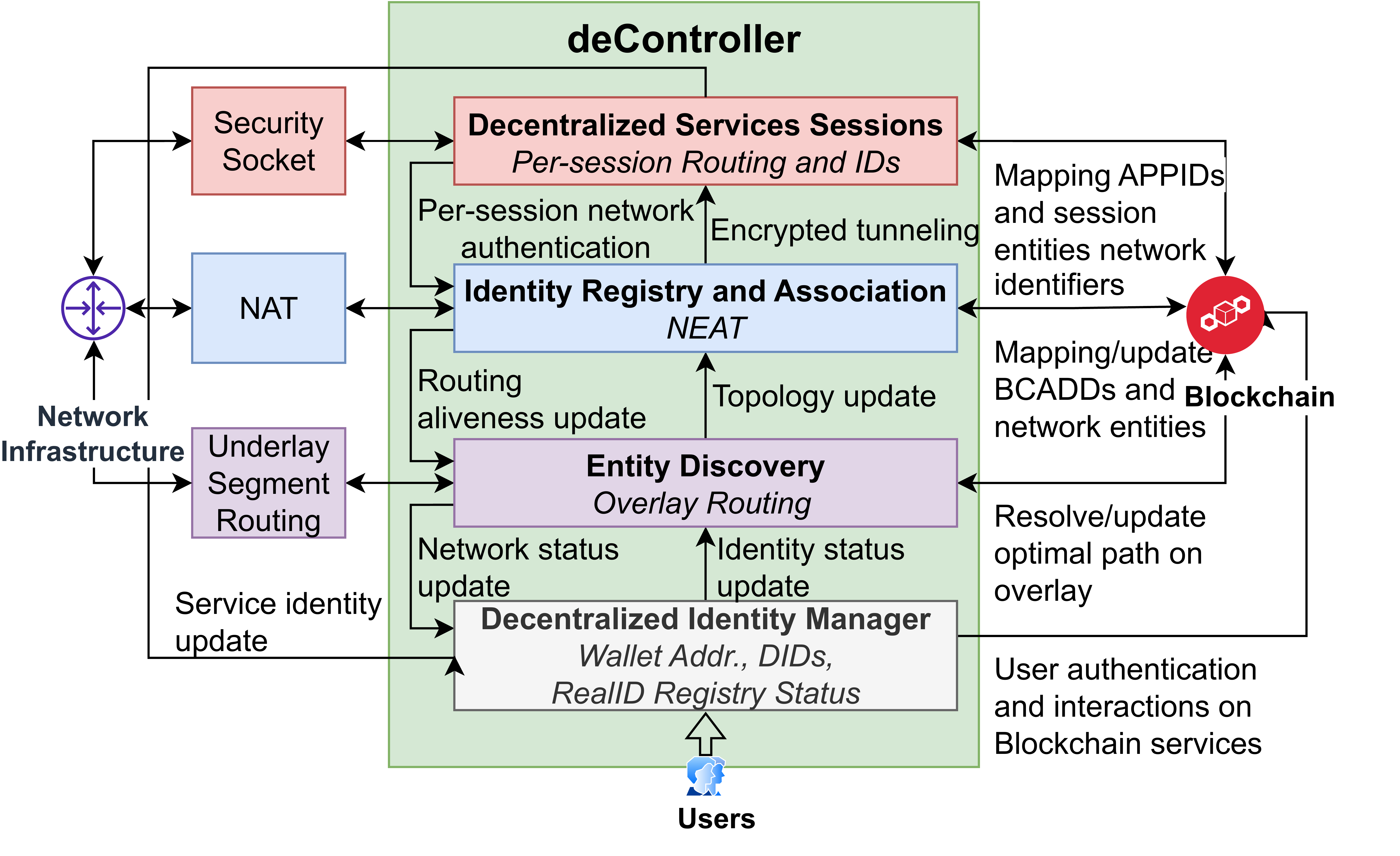}
    \caption{deController architecture.}
    \label{fig:decon}
\end{figure}

{\color{black}When a user accesses a Web3 application as shown in Fig. \ref{fig:decon}, the overlay of deController routes the encrypted application address to the corresponding smart contract deployed by the service provider. Then, a link from the user to the application can be established via the underlay of deController. After that, the user can authenticate the application and then use the smart contract to access the application via the $\sf{BCADD}$.}
The overlay network offers ultimum connectivity to decentralized users and services. However, it is still a challenge to bridge the decentralized services to the underlay physical network in a decentralized manner. In the following, we propose our deController, the nexus for the decentralized overlay and universal network underlay. 

\subsection{Identity and access with decentralized identity manager} 
One fundamental function for decentralized controllers is the authentication, including identity and service access. As discussed in Fig. \ref{fig:novelty} and Fig. \ref{fig:identity}, we introduce a hierarchy and decentralized identity management infrastructure, where ${\sf{RealID}}$, ${\sf{BCADD}}$ and ${\sf{APPID}}$ are used to achieve authentication in a privacy-preserving way.

With these $\sf{BCADDs}$, one critical innovation for Web3 architecture is to enable access with $\sf{BCADDs}$, which should be achieved in the overlay network with the help of an embedded deController, shown in Fig. \ref{fig:decon}. {\color{black}Users can access decentralized services starting from bottom to top. The left part contains network functions. Meanwhile, the smart contracts are shown on the right for user mobility and identity association updates with minor status changes recorded on the blockchain. Therefore, the blockchain is intended for small data such as identity associations and topological updates.} Specifically, the controller acts as the agency for entities to interact with the blockchain and relay the information to entities who may not support blockchain access, thus building the encrypted tunnel between two entities. Hence, the native interpretation of encrypted identities can significantly improve the security, integrity and scalability of Web3 services while pushing the boundary of decentralization towards communication infrastructures.

\subsection{Network and application integration: entity discovery}
Another key function that the deController in the overlay network should perform is the network segment routing for data delivery. In the decentralized architecture, there is no central controller to determine and update the routing table for the whole network. Therefore, deControllers determine the routing for users without using conventional network addresses. In our proposed overlay network, shown in Fig. \ref{fig:decon}, deControllers can rely on the blockchain network to perform routing optimization. Specifically, each access node is identified by its $\sf{BCADD}$ or $\sf{APPID}$, and the serving blockchain access points can be bound to addresses with the topological information. Hence, finding a data transmission path for two users is equivalent to finding a path between the two associated blockchain nodes. Thereby, a logical tunnel between two users is established with users' $\sf{BCADD}$ or $\sf{APPID}$, and further encrypted by the keys exchanged between two blockchain access points. Furthermore, the blockchain network can be mapped into multiple segments of the network, and each blockchain segment represents the overlay access point of the nearby network. The global routing topology will be collected from all blockchain routing nodes to find the routing path among different segments.

\subsection{Identity association with encrypted address translation}
Ledger records contain the information needed for routing and switching, which are essential to the self-claimed identities from clients and their current addresses' bindings. They together make up the identity registry and association services offered by deController in Fig. \ref{fig:decon}. In the case of switching, the local record utilizes the bound network interface of the entity's $\sf{BCADD}$. By having the $\sf{BCADD}$ as the pointer, the endpoint router can perform NEAT {\color{black}(an address lookup protocol based upon hash table and bloom filter)} to steer the traffic between any entities tagged within the $\sf{BCADD}$ and the connected interfaces.

\subsection{Decentralized services sessions}
As one entity can be identified with an $\sf{BCADD}$, per-session routing for each service entity can also be considered, while the traffic can be steered using the $\sf{BCADDs}$, as indicated on the top of Fig. \ref{fig:decon}. During the per-session routing, mutual authentications are performed in every handshake between two encrypted identities via the required security socket layers. Meanwhile, the subsequent service status is updated by the identity manager, who keeps tracking the service quality, aliveness and most importantly, the service identity.

\section{Conclusion}

In this paper, we propose deController, a perspective of Web3 architecture for future decentralized Web3 infrastructures, consisting of overlay and underlay to catalyze more free and fair web access for people. The functions of deController are illustrated in a top-down sculpture of Web3 architecture with the considerations of concealed identity, security and privacy, and law. The term Web3 shall also enable not only the decentralization of giant Internet companies, but also the decentralization from the de-facto centralized infrastructure controller. 
Our solution proposed in this paradigm can be a potential starting point for the real Web3 infrastructure investment, which allows the true ownership of Web3 beyond the content.

\bibliographystyle{IEEEtran}
\bibliography{references}

\begin{thebibliography}{1}
\providecommand{\url}[1]{#1}
\csname url@samestyle\endcsname
\providecommand{\newblock}{\relax}
\providecommand{\bibinfo}[2]{#2}
\providecommand{\BIBentrySTDinterwordspacing}{\spaceskip=0pt\relax}
\providecommand{\BIBentryALTinterwordstretchfactor}{4}
\providecommand{\BIBentryALTinterwordspacing}{\spaceskip=\fontdimen2\font plus
\BIBentryALTinterwordstretchfactor\fontdimen3\font minus
  \fontdimen4\font\relax}
\providecommand{\BIBforeignlanguage}[2]{{%
\expandafter\ifx\csname l@#1\endcsname\relax
\typeout{** WARNING: IEEEtran.bst: No hyphenation pattern has been}%
\typeout{** loaded for the language `#1'. Using the pattern for}%
\typeout{** the default language instead.}%
\else
\language=\csname l@#1\endcsname
\fi
#2}}
\providecommand{\BIBdecl}{\relax}
\BIBdecl

\bibitem{dao2dao}
\BIBentryALTinterwordspacing
BlockScience, ``{Exploring DAO2DAO Collaboration Mechanisms},'' 2021. [Online].
  Available:
  \url{https://medium.com/primedao/exploring-dao2dao-collaboration-mechanisms-c37218a17a21}
\BIBentrySTDinterwordspacing

\bibitem{Xu2021beran}
\BIBentryALTinterwordspacing
H.~Xu, L.~Zhang, Y.~Sun, and C.-L. I, ``{BE-RAN: Blockchain-enabled Open RAN
  with Decentralized Identity Management and Privacy-Preserving
  Communication},'' jan 2021. [Online]. Available:
  \url{http://arxiv.org/abs/2101.10856}
\BIBentrySTDinterwordspacing

\bibitem{bitansky2020verifiable}
N.~Bitansky, ``Verifiable random functions from non-interactive
  witness-indistinguishable proofs,'' \emph{Journal of Cryptology}, vol.~33,
  no.~2, pp. 459--493, 2020.

\bibitem{zeestar}
S.~Steffen, B.~Bichsel, R.~Baumgartner, and M.~Vechev, ``Zeestar: Private smart
  contracts by homomorphic encryption and zero-knowledge proofs,'' in
  \emph{2022 IEEE Symposium on Security and Privacy (SP)}, 2022, pp. 179--197.

\bibitem{33.220}
3rd Generation Partnership Project; Technical Specification Group~Services and
  S.~Aspects, ``Generic authentication architecture (gaa), generic
  bootstrapping architecture (gba), release 17, v17.3.0,'' \emph{3GPP Standard
  TS 33.220}, Jun. 2022.

\bibitem{Xu2022metaverse}
H.~Xu, Z.~Li, Z.~Li, X.~Zhang, Y.~Sun, and L.~Zhang, ``{Metaverse Native
  Communication : A Blockchain and Spectrum Prospective},'' in \emph{2022
  International Conference on Communications}, 2022.

\bibitem{Finck2019}
M.~Finck, ``Blockchain and the general data protection regulation: Can
  distributed ledgers be squared with blockchain and the general data
  protection regulation law?'' \emph{European Parliamentary Research Service},
  2019.

\end{thebibliography}
\begin{IEEEbiographynophoto}{Hao Xu}
received the PhD in Electrical Engineering from University of Glasgow. He is going to be with Wireless Network Research Department, Huawei Technologies (UK). His research interests cover wireless communication, wireless blockchain consensus and blockchain-enabled radio access network.
\end{IEEEbiographynophoto}

\begin{IEEEbiographynophoto}{Yunqing Sun}
 is currently working toward a Ph.D. degree in Computer Science, Northwestern University, US. Her research interests mainly focus on security and privacy. She is working on multi-party computation and oblivious transfer.
\end{IEEEbiographynophoto}
\begin{IEEEbiographynophoto}{Zihao Li }
is currently a PhD candidate in CREATe Center, School of Law, University of Glasgow. His research interests concentrate on the relationship between law, data and information technology.
\end{IEEEbiographynophoto}
\begin{IEEEbiographynophoto}{Yao Sun}
is currently a Lecturer with the James Watt School of Engineering, the University of Glasgow, Glasgow, U.K. His research interests include intelligent wireless networking, network slicing, blockchain system, Internet of Things and resource management in mobile networks. 
\end{IEEEbiographynophoto}
\begin{IEEEbiographynophoto}{Lei Zhang (Senior Member, IEEE)}
is a Professor at the University of Glasgow. He has academia and industry combined research experience on wireless communications and networks, and distributed systems for IoT, blockchain, autonomous systems. He is the founding Chair of IEEE Special Interest Group on Wireless Blockchain Networks in Cognitive Networks Technical Committee. 
\end{IEEEbiographynophoto}
\begin{IEEEbiographynophoto}{Xiaoshuai Zhang} is currently a Research Associate in James Watt School of Engineering of University of Glasgow. He received his Ph.D. from Queen Mary University of London. His current research interests include blockchain, distributed consensus, applied cryptography, privacy preservation and IoT.
\end{IEEEbiographynophoto}
\end{document}